\documentclass[preprint,onecolumn,aps,prd,groupedaddress,nofootinbib,showpacs]{revtex4}
\usepackage{amssymb}
\usepackage{graphicx}
\graphicspath{{fig/}{dia/}{plot/}}
\DeclareGraphicsExtensions{.eps}
\usepackage{amsmath}
\usepackage{bm}
\usepackage[hypertex]{hyperref}
\newcommand{\jpsi}{J/\psi}

\def\gev{\mathrm{~GeV}}
\def\tev{\mathrm{~TeV}}
\def\br{\mathrm{Br}}
\def\muL{\mu_{\Lambda}}

\def\be{\begin{equation}}
\def\ee{\end{equation}}
\def\bea{\begin{eqnarray}}
\def\eea{\end{eqnarray}}
\def\NO{\nonumber}
\def\vv{\overrightarrow{\textbf{v}}}
\def\EO{\Lambda}

\def\a{\alpha}

\def\s{\sigma}

\def\moh{{\langle\mathcal{O}^{H}_n\rangle}}

\def\sa{{\bigl.^1\hspace{-1mm}S^{[8]}_0}}
\def\sb{{\bigl.^3\hspace{-1mm}S^{[8]}_1}}
\def\pj{{\bigl.^3\hspace{-1mm}P^{[8]}_J}}
\newcommand{\up}[1]{\Upsilon(#1S)}
\newcommand{\xb}[2]{\chi_{b#1}(#2P)}
\newcommand{\mo}[5]{\langle\mathcal{O}^{#1}(\bigl.^#2\hspace{-1mm}#3_#4^{[#5]})\rangle}
\begin{document}
\title{\mbox{}\\[10pt] $\Upsilon(1S)$ prompt production at the Tevatron and LHC in nonrelativistic QCD}
\author{Kai Wang$~^{(a)}$, Yan-Qing Ma$~^{(a)}$, and Kuang-Ta Chao$~^{(a,b)}$}
\affiliation{ {\footnotesize (a)~Department of Physics and State Key
    Laboratory of Nuclear Physics and Technology, Peking University,
    Beijing 100871, China}\\
  {\footnotesize (b)~Center for High Energy Physics, Peking
    University, Beijing 100871, China}}
\begin{abstract}
With nonrelativistic QCD factorization, we calculate the
$\Upsilon(1S)$ prompt production at hadron colliders at
next-to-leading order in $\alpha_s$. In addition to the
color-singlet contribution, color-octet channels (especially
the P-wave channel) up to $O(v^4)$  are all considered. Aside from
direct production, the feed-down contributions from higher excited
S-wave and P-wave $b\bar b$ states to $\Upsilon(1S)$ production are
also included. We use the potential model estimates as input for
color-singlet long-distance matrix elements (LDMEs). While for
color-octet contributions, we find they can be approximately described by
three LDMEs: $\mo{}{3}{S}{1}{8}$, $\mo{}{1}{S}{0}{8}$ and
$\mo{}{3}{P}{0}{8}$. By fitting the Tevatron data we can determine
some linear combinations of these LDMEs, and then use them to
predict $\Upsilon(1S)$ production at the LHC. Our predictions are
consistent with the new experimental data of CMS and LHCb.
\end{abstract}
\maketitle
\section{INTRODUCTION}\label{sec:introduction}

The study of heavy quarkonium production is particularly interesting
because it may provide decisive information in understanding
hadronization of heavy quarks and gluons in QCD. The most widely
accepted theory to describe heavy quarkonium production at present
is nonrelativistic QCD (NRQCD) factorization \cite{Bodwin:1994jh},
in which the production is factorized into perturbative calculable
short-distance coefficients and nonperturbative (and universal)
long-distance matrix elements (LDMEs). As short-distance
coefficients can be expanded in strong-coupling $\a_s$ and each LDME
has a definite power in $v$ (the velocity of heavy quarks in the
rest frame of heavy quarkonium), NRQCD factorization gives
predictions by double expansion in $\a_s$ and $v^2$.  NRQCD
factorization is efficient such that in principle only a finite
number of universal parameters, which can be determined by using
some known experimental data, are involved with required precision
in predicting other production processes . Although a complete proof
of factorization is still lacking, at least it holds up to
next-to-next-to-leading order in $\a_s$
\cite{Nayak:2005rw,Nayak:2005rt}.

Based on NRQCD factorization, charmonium production in hadron
colliders has been studied extensively in recent
years\cite{Artoisenet:2007xi,He:2009zzb,Li:2011yc,Campbell:2007ws,Gong:2008sn,Gong:2008hk,Gong:2008ft,Ma:2010vd,Ma:2010yw,
Butenschoen:2009zy,Ma:2010jj,Butenschoen:2012px,Chao:2012iv,Kang:2011mg}.
Specifically, for $\jpsi$ hadroproduction, it is found that all data
at large $p_T$, including both yield and polarization, can be well
described by NRQCD factorization if one chooses a large $M_0$ and a
small $M_1$ \cite{Chao:2012iv} (see also
Refs.\cite{Ma:2010yw,Ma:2010jj}). Here $M_0$ and $M_1$ are linear
combinations of related LDMEs which are defined in
Refs.\cite{Ma:2010yw,Ma:2010jj} and will also be mentioned below,
and roughly speaking, their values represent the importance of
$p_T^{-6}$ behavior and $p_T^{-4}$ behavior in $\jpsi$ production
cross sections, respectively.

There are reasons that studying bottomonium may be a more suitable
choice than charmonium to test the NRQCD factorization formalism.
First, the value of $v^2$ is smaller in bottomonium ($\approx 0.1$)
than that in charmonium ($\approx 0.3$), thus the expansion in $v^2$
should converge faster in bottomonium. Second, the mass of
bottomonium is about 3 times of that of charmonium, then asymptotic
freedom implies the convergence of $\a_{s}$ expansion is also better
in bottomonium. However, on the experiment side, the situation is
not so satisfactory as the production rates of bottomonium are much
smaller than charmonium, e.g., the cross section of $\Upsilon(1S)$
is about 2 orders of magnitude smaller than that of $\jpsi$.
Furthermore, there are more excited bottomonium states which are
below the open bottom (say $B\bar B$) threshold and can decay into
lower bottomonium states such as $\Upsilon(1S)$ with large branching
ratios and consequently contribute a substantial fraction to the
lower bottomonium inclusive production by the so-called feed-down
contributions, thus it is hard to measure the direct production from
the prompt inclusive production. In the LHC era, we expect these
disadvantages may be overcome by the higher luminosity, thus testing
NRQCD factorization by bottomonium production seems to be hopeful.

The inclusive differential cross section and polarization of
$\Upsilon$ are measured at the Tevatron
\cite{Acosta:2001gv,Abazov:2005yc,:2008za,Kuhr:2010pq,CDF:2011ag}, but the
$\Upsilon(1S)$ polarizations observed by D0 \cite{:2008za} and CDF
\cite{Kuhr:2010pq,CDF:2011ag} disagree with each other. Furthermore, both the
D0 and CDF measurements contradict the LO NRQCD prediction
\cite{Braaten:2000gw}. As argued in Refs.\cite{Ma:2010vd,Ma:2010jj},
the next-to-next-to-leading order and even higher-order contributions
\cite{Artoisenet:2008fc} may not be important, as compared with the
full next-to-leading order (NLO) QCD contributions including both color-singlet (CS) and
color-octet (CO) channels, which are essential in understanding the
$\Upsilon$ (and similarly the $\jpsi$) hadroproduction. Partial NLO
QCD contributions to $\Upsilon$ hadroproduction have been calculated
recently
\cite{Campbell:2007ws,Artoisenet:2007xi,Gong:2008hk,Gong:2010bk},
and it is found that the NLO QCD corrections of S-wave CO channels
only slightly change the transverse momentum distribution and the
polarization, while the correction of CS channel may bring on
significant enhancement to the momentum distribution and change the
polarization from transverse at LO into longitudinal at NLO. But the
NLO contributions of P-wave channels for $\Upsilon$
hadroproduction are still missing.

At the LHC, CMS has published the first run data for $\Upsilon$
production \cite{Khachatryan:2010zg}, and LHCb has also reported the
measured result \cite{upslhcb}. Thus it is timely to present a
complete NLO theoretical prediction for $\Upsilon$ production, and
compare theory with experiment. In this work, we study the
$\Upsilon(1S)$ hadroproduction in the framework of NRQCD, including
all NLO contributions and feed-down contributions.
The paper is organized as follows. We briefly introduce our
calculation in Sec. \ref{sec:method}. Then in Sec.
\ref{sec:feeddown}, we describe our method for taking into account
the feed-down contributions. (Note that the feed-down contributions
for $\Upsilon(1S)$ have not been treated seriously in all previous
theoretical works.) In Sec. \ref{sec:result}, we fit data to
determine LDMEs and then give predictions for the LHC experiment.
Finally, a summary is given in Sec. \ref{sec:summary}.

\section{NLO calculation}\label{sec:method}
The method of NLO calculation used in this work is similar to
that used in $\jpsi$ and $\chi_c$ production
\cite{Ma:2010vd,Ma:2010yw,Ma:2010jj}. For completeness, we will
sketch it in this section.

According to the NRQCD
factorization formalism, the inclusive cross section for direct
bottomonium $H$ production in hadron-hadron collisions is expressed as
\begin{equation}\begin{split}
 \label{eq:NRQCD}
d\s[pp\rightarrow H+X]&=\sum\limits_{n}d\hat{\s}[(b\bar{b})_n]\dfrac{\moh}{m_b^{2L_n}}\\
&=\sum\limits_{i,j,n}\int dx_1 dx_2 G_{i/p}G_{j/p}
\times d\hat{\s}[i+j\rightarrow (b\bar{b})_n +X]\moh,
\end{split}\end{equation}
where $p$ is either a proton or an antiproton, the indices $i, j$
run over all the partonic species, and $n$ denote the color, spin
and angular momentum ($L_n$) of the intermediate $b\bar{b}$ states.
In this work, we calculate the cross sections up to $v^4$
corrections, so that the intermediate states include $^3S^{[1]}_1$,
$^3P^{[1]}_J$, $\sa$, $\sb$ and $\pj$. Note that our definition of
CS LDMEs $\mo{H}{3}{S}{1}{1}$ and $\mo{H}{3}{P}{J}{1}$ are different
from that in Ref. \cite{Bodwin:1994jh} by a factor of $1/(2N_c)$.
The calculation proceeds with three steps: calculating the parton
level differential cross section $d\hat{\s}[i+j\rightarrow
(b\bar{b})_n +X]$, integrating over the phase space, and fitting the
LDMEs.

NLO corrections for the parton level differential cross section
include virtual corrections and real corrections. For virtual
corrections, we use {\tt FeynArts} \cite{Hahn:2000kx} to generate
Feynman diagrams and amplitudes. We then calculate these thousands
Feynman diagrams analytically using our self-written {\tt
Mathematica} code. Finally, we output the simplified expression into
{\tt C++} code. Because the infrared divergence will appear
when doing phase space integration for the real correction, we use
the two cutoff phase space slicing method \cite{Harris:2001sx} to
isolate the divergence. The contributions from the singular phase
space part are calculated analytically, while finite parts are
calculated by using the Berends-Giele off-shell recursive relations
\cite{Berends:1987me}.

In the analytical calculation we have checked that all the
ultraviolet (UV) and infrared (IR) singularities are canceled
exactly. The UV divergences are removed by renormalization. The IR
singularities arising from loop integration and phase space
integration of the real correction partially cancel each other. The
remaining IR singularities are absorbed into the proton
parton-distribution functions and the NRQCD LDMEs.

The numerical integration over the phase space is handled by our
self-written {\tt C++} codes, where we also use both {\tt QCDLoop}
\cite{Ellis:2007qk} and {\tt LoopTools} \cite{Hahn:1998yk} to
calculate the scalar functions in the virtual corrections
numerically. We verified that our results are independent of the two
cuts introduced by the phase space slicing method. The method of
fitting LDMEs will be discussed in Sec. \ref{sec:result}.

\section{TREATMENT OF FEED-DOWN CONTRIBUTION}\label{sec:feeddown}

One difficulty in predicting $\Upsilon$ production cross section is
the treatment of feed-down contribution. There are several higher
excited states that can decay into $\up{1}$ and they include:
$\up{2}$, $\up{3}$, $\xb{1}{1}$, $\xb{2}{1}$, $\xb{1}{2}$ and
$\xb{2}{2}$. Unfortunately, there are not enough data to determine
LDMEs of these higher excited states, therefore, it is hard to
predict their feed-down contributions to $\up{1}$. In fact, all
previous predictions that were based on NRQCD factorization did not
have a serious treatment of the feed-down contributions.

The key point to deal with feed-down contribution is to determine
the relation between momentum of higher excited states and momentum
of $\up{1}$. In Ref. \cite{Ma:2010yw}, we find a very good
approximation that the ratio of two momenta is inversely
proportional to the ratio of their masses. Notice that the mass
differences between these excited states and $\up{1}$ are of the
order of $m_b v^2$ and  $v^2$ is very small in bottomonium, as a
result, unlike the $\jpsi$ case, the momentum shift can be ignored
when these excited states decay into $\up{1}$. Hence the production
LDMEs can be approximately combined into 6 independent ones:
\begin{equation}\begin{split}
  \label{eq:comb}
  \mo{}{3}{S}{1}{1}=&\mo{\up{1}}{3}{S}{1}{1}
  +\sum_{n=2,3}\mo{\up{n}}{3}{S}{1}{1}\br(\up{n}\rightarrow\up{1}),\\
  \mo{}{3}{P}{1}{1}=&
  \sum_{n=1,2}\mo{\xb{1}{n}}{3}{P}{1}{1}\br(\xb{1}{n}\rightarrow\up{1}),\\
  \mo{}{3}{P}{2}{1}=&
  \sum_{n=1,2}\mo{\xb{2}{n}}{3}{P}{2}{1}\br(\xb{2}{n}\rightarrow\up{1}),\\
  \mo{}{3}{S}{1}{8}=&\mo{\up{1}}{3}{S}{1}{8}
  +\sum_{n=2,3}\mo{\up{n}}{3}{S}{1}{8}\br(\up{n}\rightarrow\up{1})\\
  &+\sum_{n=1,2}\sum_{J=1,2}\mo{\xb{J}{n}}{3}{S}{1}{8}\br(\xb{J}{n}\rightarrow\up{1}),\\
  \mo{}{1}{S}{0}{8}=&\mo{\up{1}}{1}{S}{0}{8}
  +\sum_{n=2,3}\mo{\up{n}}{1}{S}{0}{8}\br(\up{n}\rightarrow\up{1}),\\
  \mo{}{3}{P}{0}{8}=&\mo{\up{1}}{3}{P}{0}{8}
  +\sum_{n=2,3}\mo{\up{n}}{3}{P}{0}{8}\br(\up{n}\rightarrow\up{1}).
\end{split}\end{equation}
Here the $\chi_{b0}(1P,2P)$ feed-down into $\Upsilon(1S)$ is ignored
due to the smallness of the transition branching ratios.
The potential model results of wave functions and
their derivatives at the origin can be chosen
as\cite{Eichten:1995ch}
\begin{equation}\begin{split}
\lvert R_{\up{1}}(0) \rvert^{2}&= 6.477 \gev^3, \\
\lvert R_{\up{2}}(0) \rvert^{2}&= 3.234 \gev^3, \\
\lvert R_{\up{3}}(0) \rvert^{2}&= 2.474 \gev^3, \\
\lvert R_{\xb{}{1}}^{\prime}(0) \rvert^{2}&= 1.417 \gev^5, \\
\lvert R_{\xb{}{2}}^{\prime}(0) \rvert^{2}&= 1.653 \gev^5,
\end{split}\end{equation}
and the CS LDMEs can be estimated by
\begin{equation}\begin{split}
  \mo{\up{n}}{3}{S}{1}{1}&=\frac{3}{4\pi}\lvert R_{\up{n}}(0) \rvert^{2},\\
  \mo{\xb{J}{n}}{3}{P}{J}{1}&=\frac{3}{4\pi}\lvert R_{\xb{}{n}}^{\prime}(0) \rvert^{2}(2J+1).
\end{split}\end{equation}
With the PDG data of branching ratios \cite{Nakamura:2010zzi}, we
get
\begin{equation}\begin{split}
  \mo{}{3}{S}{1}{1}=& 1.81 \gev^3,\\
  \mo{}{3}{P}{1}{1}=& 0.54 \gev^5,\\
  \mo{}{3}{P}{2}{1}=& 0.62 \gev^5.
\end{split}\end{equation}
Now there leave only 3 unknown CO LDMEs: $\mo{}{3}{S}{1}{8}$,
$\mo{}{1}{S}{0}{8}$, and $\mo{}{3}{P}{0}{8}$. They will be
determined by fitting the Tevatron data
\cite{Acosta:2001gv,Abazov:2005yc}. Because they are nearly
universal (up to a correction of order $v^2$ with calculated
short-distance coefficients in the fit), the fitted results can be
used to predict $\up{1}$ production in other colliders.

\section{NUMERICAL RESULT}\label{sec:result}
The CTEQ6L1 and CTEQ6M parton-distribution
functions \cite{Whalley:2005nh} are used for LO and NLO calculations,
respectively. The bottom quark mass is set to be $m_b = 4.75 \gev$,
while the renormalization, factorization, and NRQCD scales are
$\mu_r = \mu_f = m_T$ and $\muL = m_b$, where $m_T = \sqrt{p_T^2 +
4m_b^2}$ is the $\Upsilon$ transverse mass. The center-of-mass
energies are 1.8 $\tev$, 1.96 $\tev$ and 7 $\tev$ for the Tevatron
RUN I, RUN II, and LHC, respectively.

\begin{figure*}[htbp]
\begin{tabular}{cc}
  \includegraphics[width=.45\linewidth]{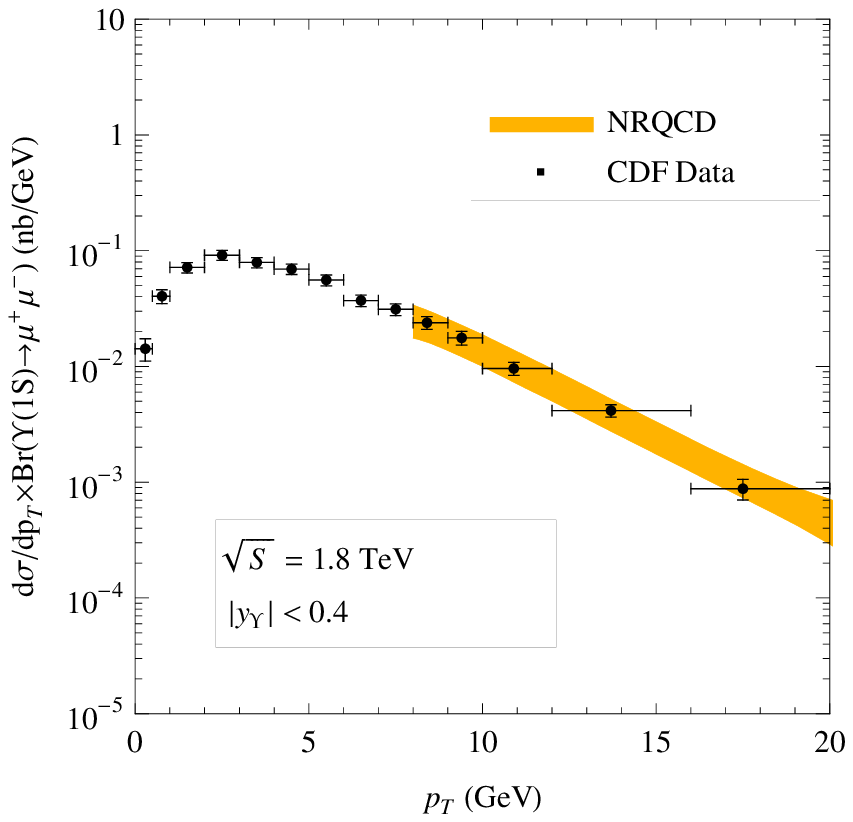} &
  \includegraphics[width=.45\linewidth]{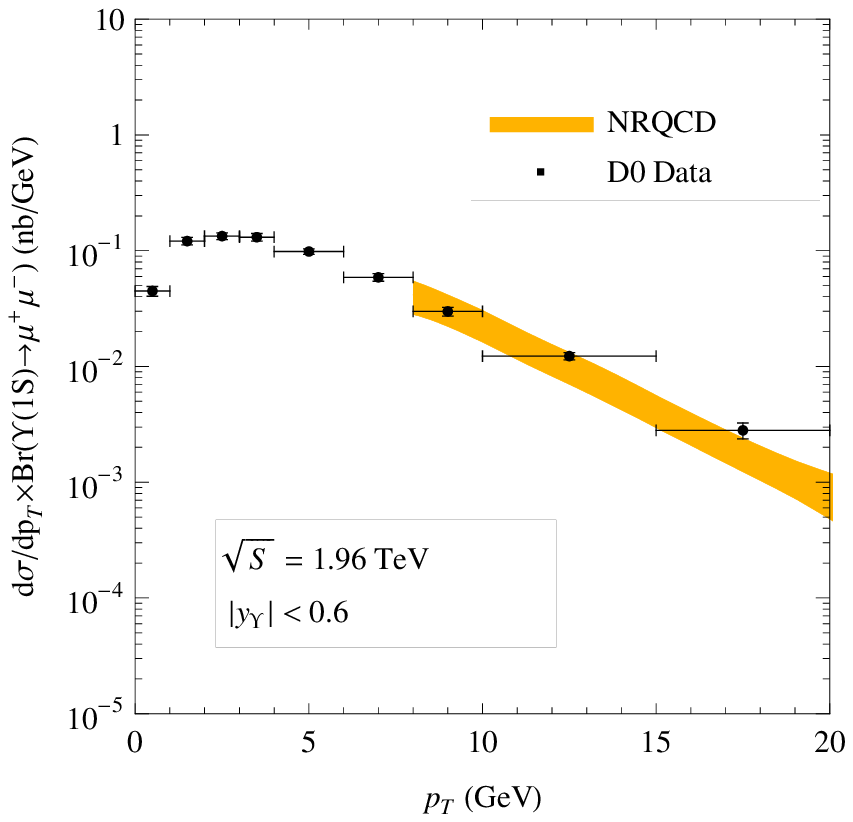} \\
  \includegraphics[width=.45\linewidth]{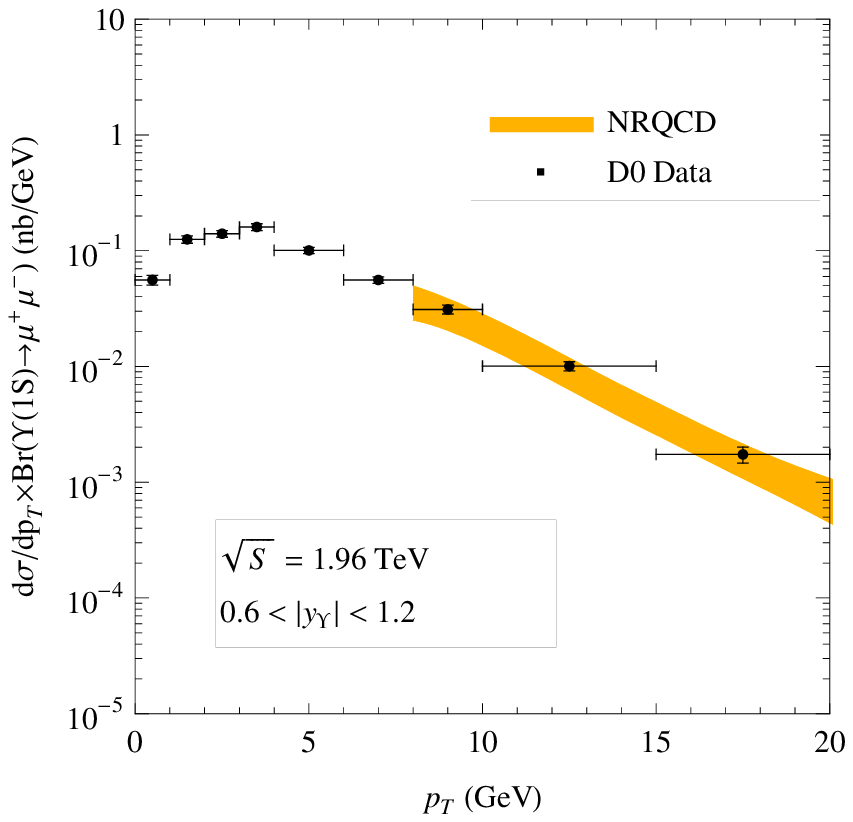} &
  \includegraphics[width=.45\linewidth]{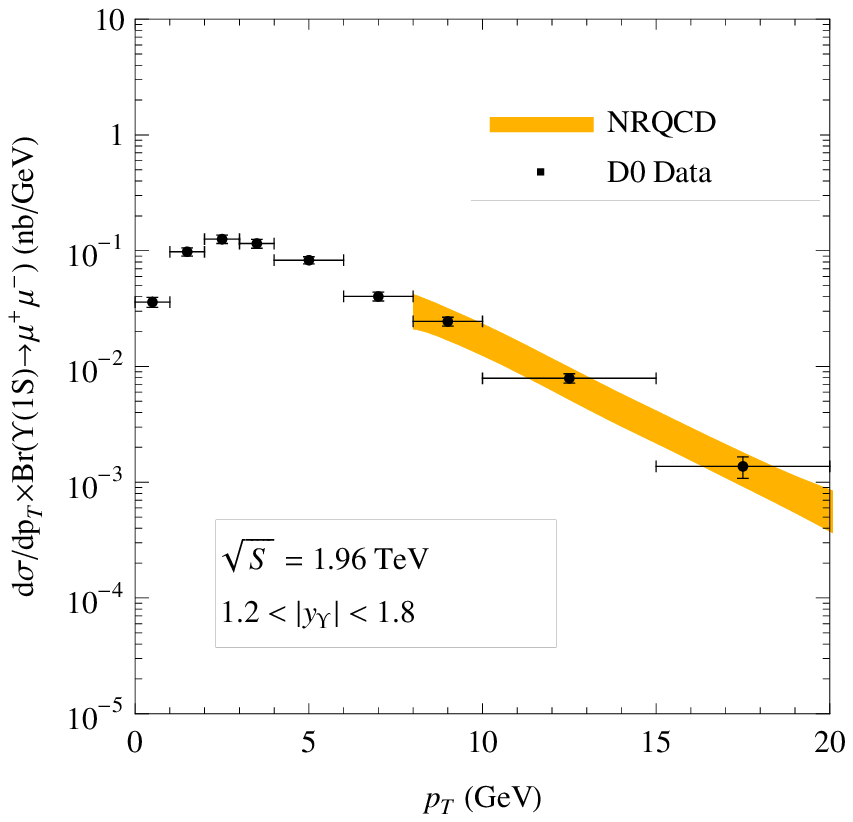}
\end{tabular}
\caption{\label{fig:tev}
  Transverse momentum
  distributions of prompt $\Upsilon(1S)$ production cross sections at the Tevatron.
  The CDF data are taken from Ref. \cite{Acosta:2001gv}.
  The D0 data are taken from Ref. \cite{Abazov:2005yc}.
}
\end{figure*}
In the fit we introduce a $p_T^{cut}$, and the Tevatron data
\cite{Acosta:2001gv,Abazov:2005yc} in Fig. \ref{fig:tev} with $p_{T}
> p_T^{cut}$ are used to fit the 3 unknown CO LDMEs:
$\mo{}{3}{S}{1}{8}$, $\mo{}{1}{S}{0}{8}$ and $\mo{}{3}{P}{0}{8}$.
The reason for discarding the low $p_T$ data is that these data are
far from the large $p_T$ region ($\frac{m_b}{p_T}\ll 1$), and may
also be affected by nonperturbative effects, which can not be
described by our fixed order perturbative calculation. If we choose
too large $p_T^{cut}$, there are no enough data to determine the CO
LDMEs, so we choose $p_T^{cut}$ = 8 GeV. Anyway, by varying
$p_T^{cut}$ from 7 GeV to 9 GeV, we find the determined CO LDMEs in
Eq.(12) are roughly consistent within errors. As discussed in Refs.
\cite{Ma:2010yw,Ma:2010jj}, by fitting large $p_T$ data at the
Tevatron one can only constrain two linear combinations of LDMEs
that have $p_T^{-4}$ and $p_T^{-6}$ behaviors at parton level,
respectively. Thus, to have a constrained fit, the short-distance
coefficient of P-wave channel is decomposed into linear combination
of that of two S-wave channels in Ref. \cite{Ma:2010yw}, and as a
result, one needs to fit only two linear combinations of LDMEs
($M_0$ and $M_1$). However, for $\Upsilon$ production we find the
decomposition of P-wave channel is good only for $p_T > 15 \gev$,
thus, to fit the Tevatron data with $p_{T} \gtrsim  8 \gev$, we
cannot decompose the P-wave channel. Therefore, we fit the three
LDMEs using a similar method described in Ref. \cite{Ma:2010jj}.

Define
 \bea \label{fit3value}
O_1 \equiv \mo{}{1}{S}{0}{8},\NO\\
O_2 \equiv \mo{}{3}{S}{1}{8},\\
O_3 \equiv \frac{\mo{}{3}{P}{0}{8}}{m_b^2},\NO
 \eea
and the correlation matrix C
 \bea C_{ij}^{-1} = \frac{1}{2} \frac{d^2 \chi^2}{d O_i d O_j}.
 \eea
By minimizing $\chi^2$ we have
 \bea
C = \left(
  \begin{array}{ccc}
    0.24 & -0.024 & -0.54 \\
    -0.024 & 0.0025 & 0.054 \\
    -0.54 & 0.054 & 1.21 \\
  \end{array}
\right).
 \eea
The eigenvalues $\lambda_i$ with corresponding eigenvectors $\vv_i$
of C are
 \bea \label{eigen}
 &\lambda_1 = 1.5               ,& \vv_1 = (-0.41, 0.040, 0.91)\NO \\
 &\lambda_2 = 3.5 \times 10^{-4},& \vv_2 = (0.79, -0.48, 0.38) \\
 &\lambda_3 = 1.3 \times 10^{-5},& \vv_3 = (0.46, 0.87, 0.17).\NO
 \eea
The LDMEs corresponding to the eigenvectors are
 \bea \label{DeLambda}
\left(
  \begin{array}{c}
    \EO_1 \\
    \EO_2 \\
    \EO_3 \\
  \end{array}
\right)
 = V \left(
  \begin{array}{c}
    O_1 \\
    O_2 \\
    O_3 \\
  \end{array}
\right),
 \eea
where we denote matrix
 \bea
 V = \left(
  \begin{array}{c}
    \vv_1 \\
    \vv_2 \\
    \vv_3 \\
  \end{array}
\right).
 \eea
Inserting Eqs.(\ref{fit3value}) and (\ref{eigen}) into
Eq.(\ref{DeLambda}), we have
 \bea \label{Lambda}
\EO_1 &=& -274 \times10^{-2}\gev^3~~(\pm44\%),\NO \\
\EO_2 &=& 6.04 \times10^{-2}\gev^3~~(\pm31\%),  \\
\EO_3 &=& 10.5 \times10^{-2}\gev^3~~(\pm3.4\%). \NO
 \eea
In this way, the three CO LDMEs are expressed in terms of their
linear combinations $\EO_i$, which correspond to the eigenvectors of
the correlation matrix. As the Tevatron data are not sensitive to
the value of $\Lambda_1$ in our fit, there is a large range value of
$\Lambda_1$ that can satisfy the data, and its determined value in
the fit is just randomly chosen from this range. If the range is
much larger than the physical value of $\Lambda_1$, there will be a
high possibility that the absolute value of its fitted value is much
larger than its physical value. Assuming the physical value of
$\Lambda_1$, $\Lambda_2$ and $\Lambda_3$ are of the same order, the
random choice implies the absolute value of the fitted value of
$\Lambda_1$ will be much larger than $\Lambda_2$ and $\Lambda_3$,
which is the case in our fit. Nevertheless, to change $\EO_1$ to be
the same order as $\EO_2$ and $\EO_3$, one needs more than two
$\sigma$ shift. It implies results in Eq.\eqref{Lambda} may
underestimate the error of $\EO_1$.

\begin{figure}[htbp]
\begin{tabular}{c}
  \includegraphics[width=.45\linewidth]{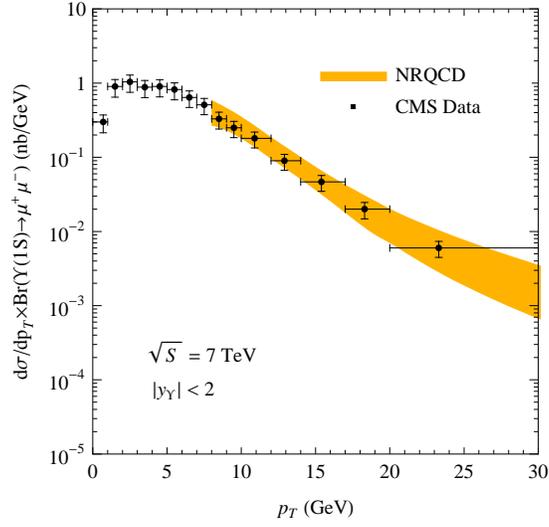} \\
  \includegraphics[width=.45\linewidth]{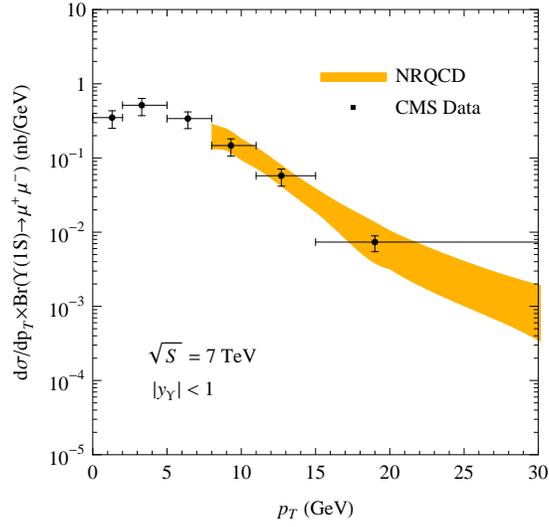} \\
  \includegraphics[width=.45\linewidth]{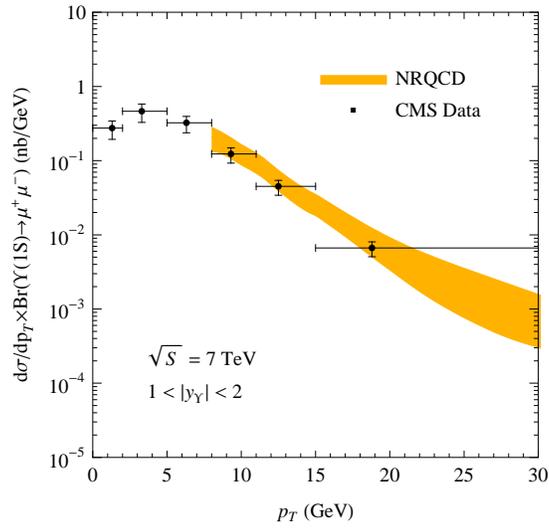}
\end{tabular}
\caption{\label{fig:cms}
  Transverse momentum
  distributions of prompt $\Upsilon(1S)$ production cross sections at the LHC.
  The CMS data are taken from Ref.\cite{Khachatryan:2010zg}.
}
\end{figure}

\begin{figure*}[htbp]
\begin{tabular}{cc}
  \includegraphics[width=.45\linewidth]{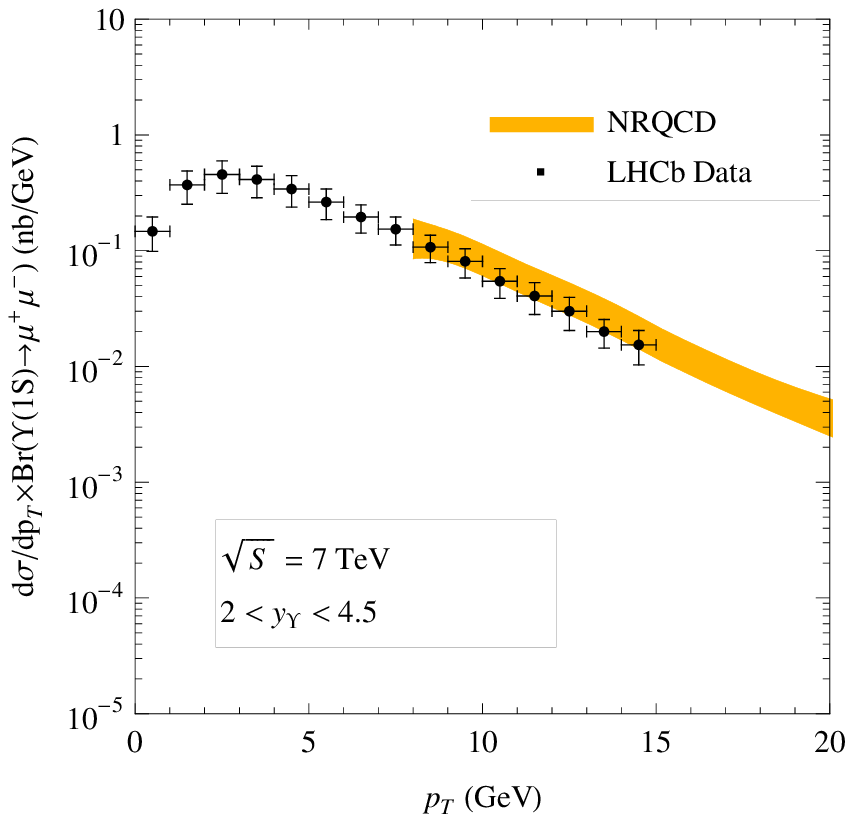} &
  \includegraphics[width=.45\linewidth]{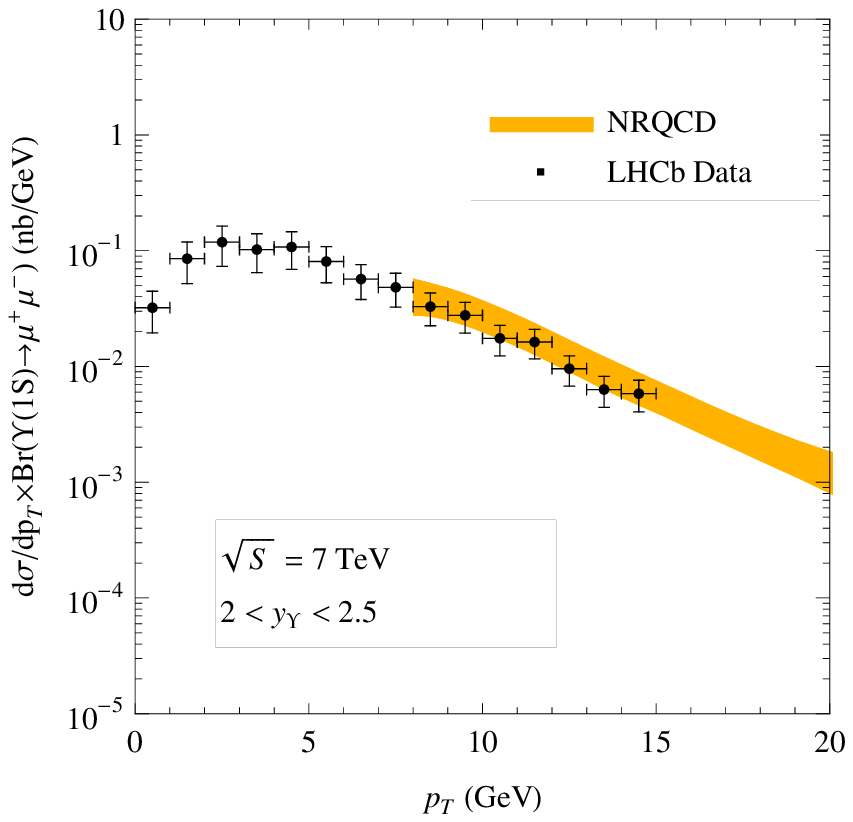} \\
  \includegraphics[width=.45\linewidth]{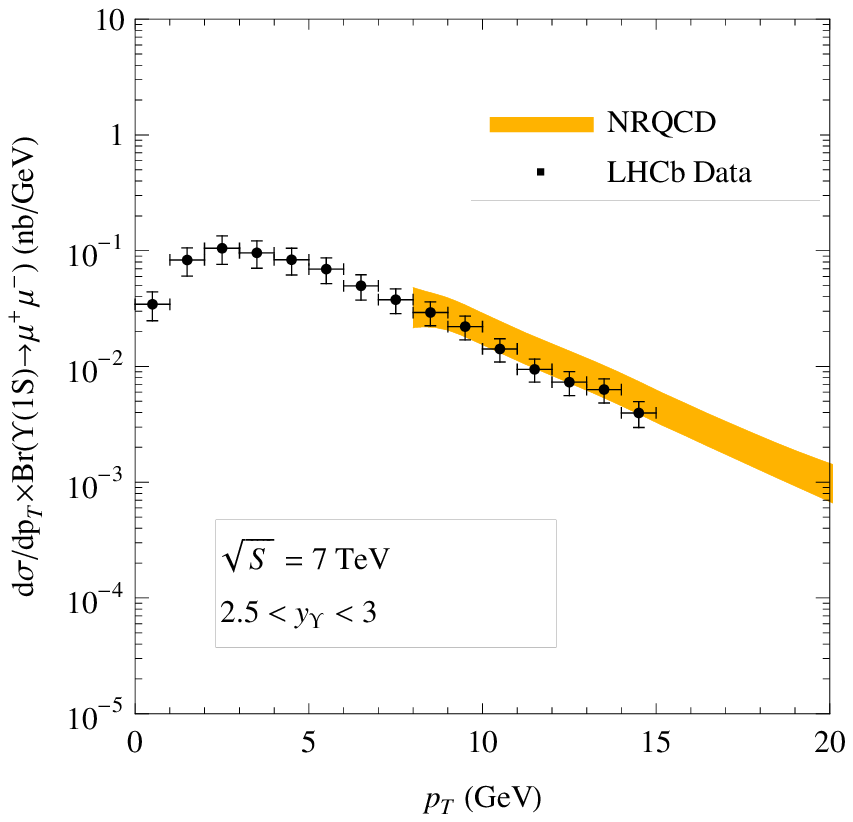} &
  \includegraphics[width=.45\linewidth]{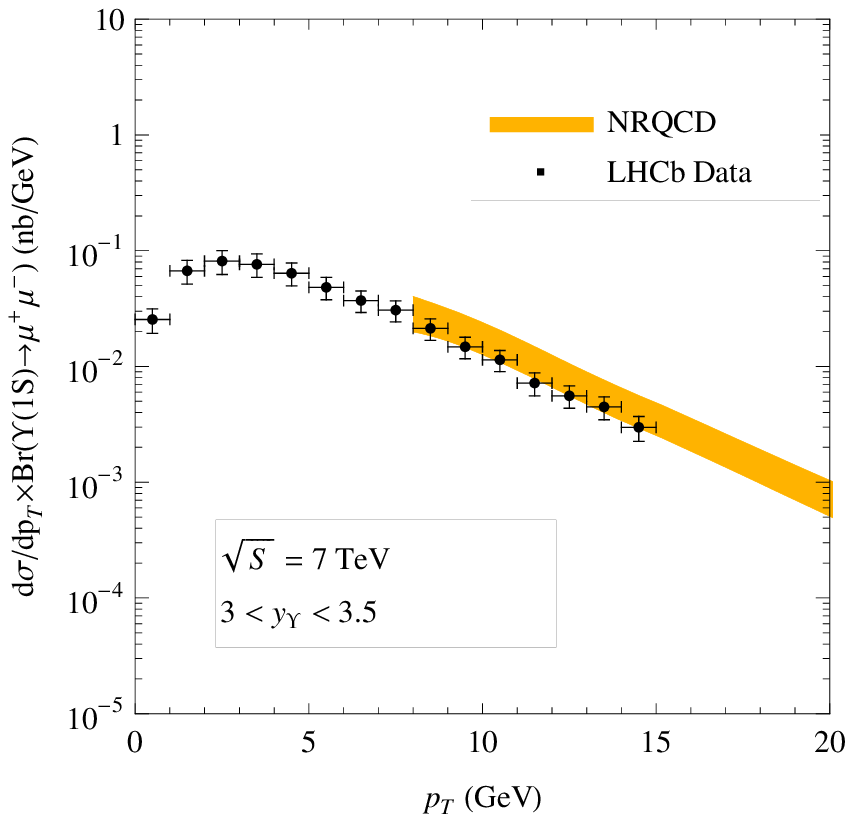} \\
  \includegraphics[width=.45\linewidth]{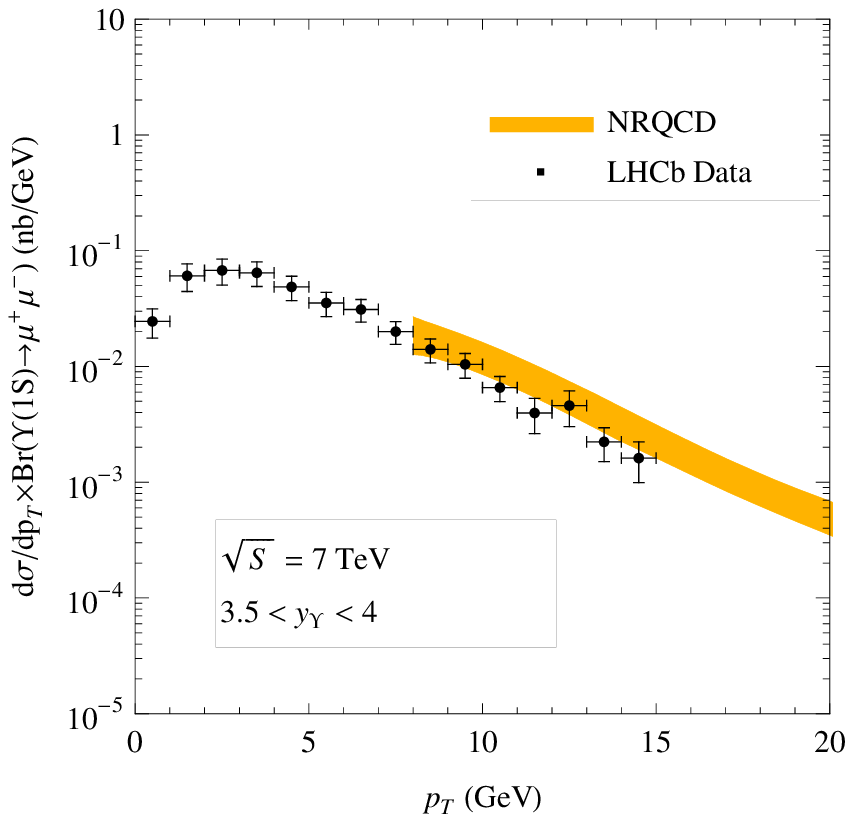} &
  \includegraphics[width=.45\linewidth]{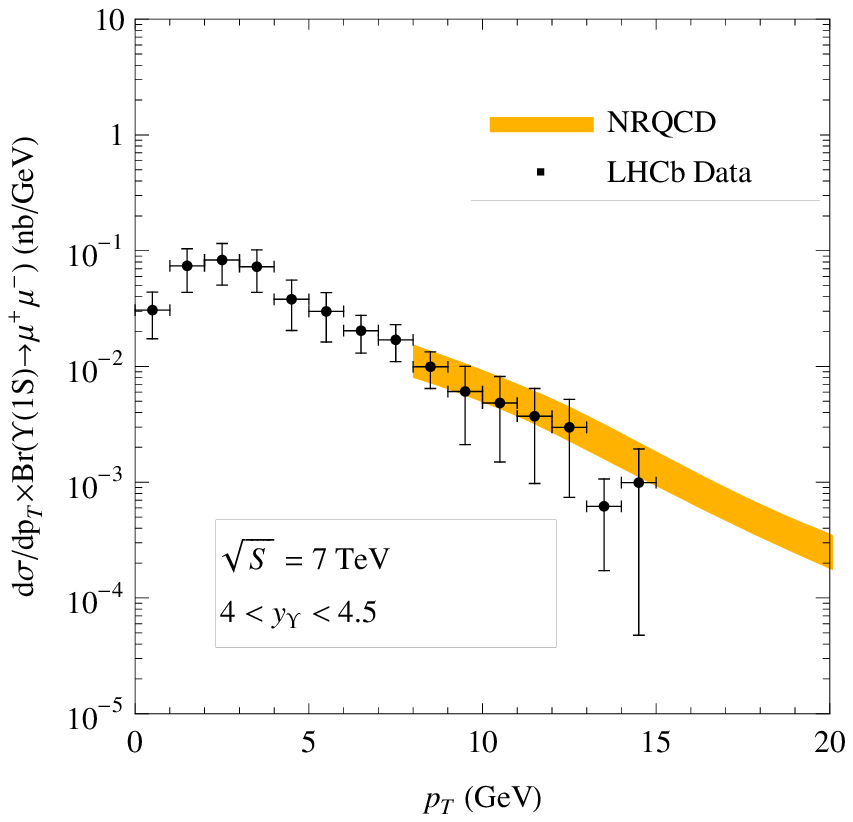}
\end{tabular}
\caption{\label{fig:lhcb}
  Transverse momentum
  distributions of prompt $\Upsilon(1S)$ production cross sections at the LHC.
  The LHCb data are taken from Ref.\cite{upslhcb}.
}
\end{figure*}

Values of $\EO_i$ contain main result in our fit. To use them to
predict $\up{1}$ production in other experiment, we express the
differential cross section as
 \bea \label{dseq}
d\s=\sum\limits_{i=1}^{3}d\hat{\s}_i O_i = \sum\limits_{i=1}^{3}a_i
\EO_i, ~~~\text{with}~
\overrightarrow{\textbf{a}}=\overrightarrow{d\hat{\s}}~ V^{-1},
 \eea
where $d\hat{\s}_i$ denote corresponding short-distance
coefficients. In this form, the errors induced by $\EO_i$ can be
easily taken into consideration for they are independent. Based on
Eq. \eqref{dseq}, our predictions for CMS and LHCb are plotted in
Fig. \ref{fig:cms} and Fig. \ref{fig:lhcb}, respectively, where CMS
and LHCb data are taken from Refs.
\cite{Khachatryan:2010zg,upslhcb}. The uncertainties of the curves
concern the renormalization scale dependence in the calculation and
the errors from $\EO_i$. We treat these two types of uncertainties
as independent ones. From these figures, we can see that our
predictions are consistent with the LHC experimental data, which is
an explicit demonstration of the universality of LDMEs defined in
Eq. \eqref{eq:comb}.


\section{SUMMARY}\label{sec:summary}

In summary, we calculate the complete NLO corrections for the
$\Upsilon(1S)$ production at hadron colliders up to
$\mathcal{O}(\a_s^4v^4)$. Ignoring corrections of higher-orders in
$v^2$, we combine the production LDMEs of $\Upsilon(1S)$ and other
excited states into 3 color-singlet LDMEs and 3 color-octet LDMEs.
These 6 LDMEs are approximately universal and they include almost
all feed-down contributions to $\Upsilon(1S)$ production. The CS
LDMEs are estimated by using potential model results, while the CO
LDMEs are determined by fitting the Tevatron data. Then we find our
predictions well coincide with the new experimental data at the LHC.
Our work may provide a new test for the universality of LDMEs in
$\Upsilon(1S)$ hadroproduction.

To have a comprehensive understanding of $\Upsilon(1S)$
hadroproduction, it is certainly important to also compare the
theoretical result with the polarization data for $\Upsilon(1S)$, we
leave it as a further study. Encouraged by the result of $\jpsi$
polarization \cite{Chao:2012iv}, where we find the $\jpsi$
polarization and yield can be consistently explained by two well
constrained CO LDMEs ($M_0$ and $M_1$), a good description for the
$\Upsilon(1S)$ data including yield and polarization seems to be
promising. However, note that the values of two well constrained CO
LDMEs in $\jpsi$ production are significantly different
\cite{Ma:2010yw,Ma:2010jj}, while for $\Upsilon(1S)$ production we
find from Eq.\eqref{Lambda} that $\EO_2$ and $\EO_3$ are of the same
order. Another complexity concerns the influence of big feed-down
contributions on $\Upsilon(1S)$ polarization. Therefore, a full
understanding of $\Upsilon(1S)$ production including both yield and
polarization may provide important information in addition to the
study of $\jpsi$ production. On the experiment side, because the
bottom quark is heavy: $m_b\approx$ 5 GeV, to test the large $p_T$
($\frac{m_b}{p_T}\ll 1$) behavior one needs to measure the cross
sections and polarizations at $p_T$ as large as, say 30 GeV and even
larger, with higher statistics, and to separate the higher excited
$b\bar b$ production from the $\Upsilon(1S)$ production. This is a
hard task for experiment, and we hope it can be fulfilled at the LHC
in the near future. Then we can make more thorough comparison
between theory and experiment, and provide a further test of NRQCD
factorization.

\begin{acknowledgments}
This work was supported by
the National Natural Science Foundation of China under Grant
Nos. 10721063,
11021092, 11075002 and the Ministry of Science and Technology of
China under Grant
No.2009CB825200.
\end{acknowledgments}
\providecommand{\href}[2]{#2}\begingroup\raggedright\endgroup

\end{document}